\documentclass[twocolumn,showpacs,preprintnumbers,amsmath,amssymb]{revtex4}
\usepackage{graphicx}% Include figure files
\usepackage{dcolumn}% Align table columns on decimal point
\usepackage{bm}% bold math
\usepackage[T1]{fontenc}
\usepackage{color}
\usepackage{tikz}
\usepackage{booktabs}
\usepackage{sistyle}
\usepackage{multirow}
\usepackage[english]{babel}
\usepackage{booktabs}
\usepackage[latin1]{inputenc}				  %schöne Tabellen

\newcommand{\D}{\textit{D}~mode}

\begin{document}

\preprint{}

\title{Resonance behavior of the defect-induced Raman mode of single-chirality enriched carbon nanotubes}% Force line breaks with \\

\author{Jan Laudenbach$^\textsf{\bfseries 1}$}
\email{jan.laudenbach@gmx.de}
\author{Frank Hennrich$^\textsf{\bfseries 2}$}
\author{Hagen Telg$^\textsf{\bfseries 3}$}
\author{Manfred Kappes$^\textsf{\bfseries 4}$}
\author{Janina Maultzsch$^\textsf{\bfseries 1}$}

\affiliation{$^1$ Institut f\"ur Festk\"orperphysik, Technische Universit\"at Berlin, Hardenbergstr. 36, D-10623 Berlin, Germany}
\affiliation{$^2$ KIT, Institut für Nanotechnologie, Hermann-von-Helmholtz-Platz 1, D-76344 Eggenstein-Leopoldshafen, Germany}
\affiliation{$^3$ CINT, Los Alamos National Laboratory, Los Alamos, New Mexico 87545, United States}
\affiliation{$^4$ KIT, Institut für Nanotechnologie, Kaiserstr. 12, D-76131 Karlsruhe, Germany}

\begin{abstract}
We present a resonance Raman study of the disorder-induced \D\ in a sample highly enriched with semiconducting (9,7) single-walled carbon nanotubes in the excitation energy range of $\num{1.49}-\SI{2.05}{eV}$. The intensity of the \D\ shows a resonance behavior near the optical transition of the (9,7) tube. The well-known dispersion of the \textit{D}-mode frequency, on the other hand, is not observed at the resonance, but only above a certain excitation energy. We explain our results by numerical simulations of the \textit{D}-mode spectra. 
\end{abstract}

\maketitle

\section{Introduction}
Raman spectroscopy is a powerful characterization tool for carbon nanotubes and graphene, due to its high sensitivity to sp$^2$-hybridized carbon \cite{reich2008carbon,Thomsen2007b,Jorio2008}. In particular, the disorder-induced \textit{D} mode and the related two-phonon \textit{2D} (\textit{G'}) mode are indicative of sp$^2$-hybridized carbon, as they stem from a breathing-like vibration of the carbon hexagons \cite{Ferrari2000} in a double-resonant Raman process (dRR) \cite{Thomsen2000}. Because of their origin the \textit{D} and \textit{2D} modes are most sensitive to the electronic band structure as well. This, among others, leads to the spectroscopic distinction between single-layer and few-layer graphene \cite{Ferrari2006}.
\\                                   
The disorder-induced \textit{D} mode, in addition, depends on the concentration of structural defects in carbon nanotubes and other graphitic materials \cite{Cancado2011}; its intensity is therefore taken as evidence for, e.g.,  successful covalent functionalization \cite{Hirsch2007, Syrgiannis2008a}. In single-walled carbon nanotubes (SWCNTs), however, the intensity and resonance behavior of the \textit{D} mode have not yet been fully understood \cite{Cardenas2007, Laudenbach2012}. This is partly due to the fact that, initially, chiral-index defined SWCNT samples were not available. Moreover, most experiments on the \textit{D} mode of carbon nanotubes investigated the dispersion of the \textit{D}-mode frequency with excitation energy. They were done either in nanotube ensembles, using a broad range of excitation energies \cite{Pimenta2000,Maultzsch2001,Brown2001,Kuerti2002}, or on individual SWCNTs, using only a small number of excitation energies and assuming that only those SWCNTs contribute to the \textit{D} mode for which the excitation energy is in resonance with an optical transition \cite{Pimenta2001,SouzaFilho2002,SouzaFilho2003}. Numerical simulations, on the other hand, focused on the \textit{D}-mode dispersion of both, individual nanotubes \cite{Maultzsch2001} and nanotube ensembles \cite{Kuerti2002}, showing an intensity decrease of the \textit{D} mode away from the optical transition of the nanotube. Ref. \cite{Kuerti2002} emphasized the intensity enhancement of the \textit{D} mode directly at optical resonances (enhanced-dRR scattering), attributing the \textit{D}-mode dispersion in nanotube ensembles to excitation-energy dependent contributions of different nanotubes with different \textit{D}-mode frequencies. However, excitation-energy dependent Raman experiments following the \textit{D}-mode dispersion and intensity on chiral-index identified carbon nanotubes have not been reported to the best of our knowledge. On the other hand, knowledge about nanotube phonon dispersion \cite{Piscanec2007,Mohr2007}, optical transitions \cite{Bachilo2002,Wang2005a,Maultzsch2005b}, and chiral-index sorting and assignment \cite{Tu2009,Stuerzl2009,Telg2011,Liu2011} has improved significantly over recent years.
\\
Here we present a resonance Raman study of the \textit{D}~mode in a sample highly enriched with (9,7)~SWCNTs in the excitation energy range of $\num{1.49}-\SI{2.05}{eV}$. The intensity of the \D\ in our sample depends strongly on excitation energy and shows a resonance behavior near the optical transition of the (9,7) SWCNT. On the other hand, the well-known dispersion of the \textit{D}-mode frequency with excitation energy  starts only above a certain excitation energy, i.e. away from the resonance. Numerical simulations of the \textit{D}-mode spectra of the (9,7) SWCNT agree with both observations, supporting the interpretation of an enhanced, only weakly dispersive double-resonant Raman scattering at excitation energies close to the optical transitions in SWCNTs.

\section{Experimental setup}
SWCNTs produced by pulsed-laser vaporization were wrapped with the polymer Poly(9,9-di-n-octylfuorenyl-2,7-diyl) (POF) in Toluol. After ultracentrifugation mainly individual (9,7) SWCNTs were dissolved in this solution \cite{Stuerzl2009}. After a time of rebundling, the sample was centrifuged again to separate the (9,7)~SWCNTs from excessive POF. This was necessary to reduce the strong Raman signal of the POF in the frequency region of the \D. The obtained solution was drop casted onto a Si/SiO$_2$ (\SI{200}{nm} SiO$_2$) substrate.\\
Raman measurements were performed with a microscope setup in backscattering geometry. A titanium-sapphire (\num{1.49}~- \SI{1.77}{eV}) and a dye laser (\num{1.79}~- \SI{2.05}{eV}) were used for excitation with \SI{1}{mW} laser power. The spectra were collected using a Dilor XY800 triple monochromator. They were calibrated with a neon lamp; the intensity was normalized to CaF$_2$ for the radial-breathing mode and to diamond for the \textit{D}~mode. All measurements were performed at the same spot of the sample.  

\section{Experiments}
\subsection{Chiral-index assignment}
\begin{figure}[tb]
	\centering
		\includegraphics[width=\linewidth]{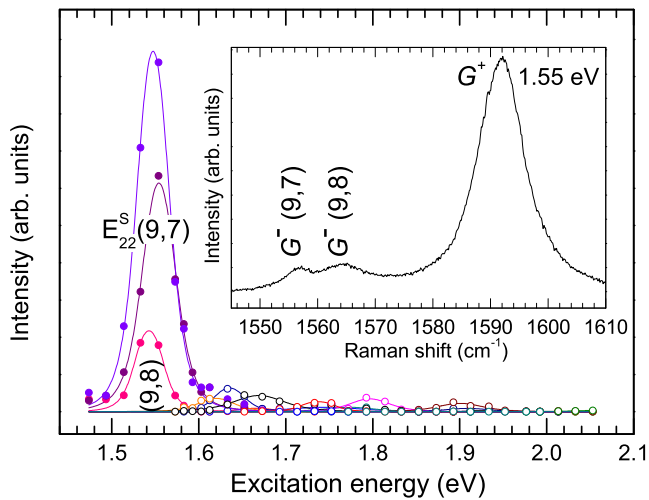}
	\caption{(Color online) RBM resonance profiles of sample enriched with (9,7)~SWCNTs with chiral-index assignment. The inset shows the Raman spectra of the \textit{G$^-$}~modes of the (9,7) and (9,8) SWCNTs at \SI{1.55}{eV} excitation energy.}
	\label{fig:rbmg}
\end{figure}

Figure~\ref{fig:rbmg} shows the resonance profiles of the radial-breathing mode (RBM) for the chiral-index assignment \cite{Maultzsch2005a}.
The profiles with the highest intensities can be assigned to the second optical transition ($E_{22}$) of semiconducting (9,7) and (9,8) SWCNTs. We found that the RBM of the (9,7)~SWCNTs consists of two contributions with slightly different frequencies ($\Delta\omega=\SI{2.5}{cm^{-1}}$) and resonance energies ($E'=\SI{1.5405}{eV}$, $E''=\SI{1.5336}{eV}$, $\Delta E=6.9\pm \SI{0.8}{meV}$). Assigning these resonance profiles to two different chiral indices seems not suitable, because of the isolated position of the (9,7)~SWCNT with respect to other SWCNTs in the Kataura plot \cite{Popov2005}. A splitting of the RBM resonance profile due to incoming and outgoing resonance would correspond to an energy difference of approximately \SI{27}{meV}, which is much higher than in our measurements. Therefore, the splitting cannot be explained up to know in detail but might be due to the enrichment process. We confirm our chiral-index assignment with help of the \textit{G}$^-$~mode (see Table \ref{tab.g-}) \cite{Telg2011}. The Raman spectrum of the \textit{G}$^-$~modes at \SI{1.55}{eV} excitation energy is shown in the inset of Fig.~\ref{fig:rbmg}.
\begin{table}[tb]																																																		
\centering												
\normalsize
\begin{tabular}{@{}cccc@{}}
\toprule 
SWCNT&\multicolumn{3}{c}{\textit{G}$^-$-mode frequency (\SI{}{cm^{-1}})}\\
\cmidrule(){1-4}
&this work&exp. Ref.~\cite{Telg2011}&theory Ref.~\cite{Telg2011}\\
(9,7)&\num{1556.7}$\pm$\num{0.8}&\num{1557.4\pm1}&\num{1558.8}\\
(9,8)&\num{1564.4}$\pm$\num{0.7}&-&\num{1561.3}\\
	\bottomrule
\end{tabular}
\caption{\textit{G}$^-$-mode frequencies of the (9,7) and (9,8) SWCNTs compared with Ref.~\cite{Telg2011}.} 
\label{tab.g-}
\end{table}
\\
Some weaker RBMs of other SWCNTs (Fig.~\ref{fig:rbmg}, open symbols) are observed above the excitation energy of \SI{1.6}{eV}, however, their intensities were at least by a factor of \num{27} lower than for the (9,7) SWCNT and by a factor of \num{3} lower than for the (9,8) SWCNT.     

\subsection{Dispersion and resonance behavior of the defect-induced Raman mode}
\begin{figure}[tb]
	\centering
		\includegraphics[width=\linewidth]{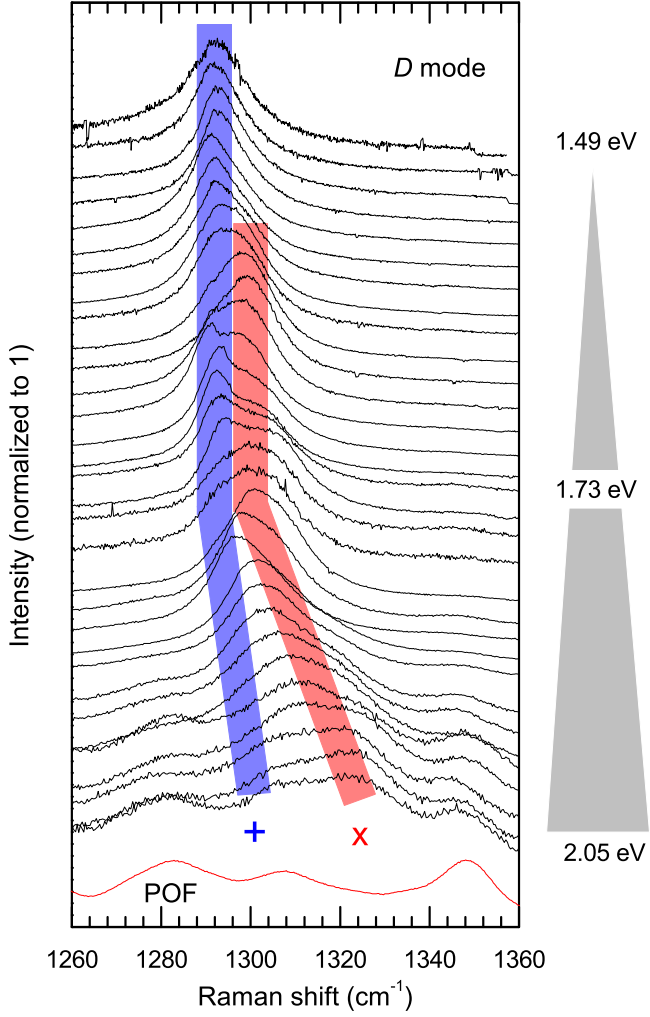}
	\caption{(Color online) Raman spectra of the \D\ of the sample enriched with (9,7)~SWCNTs for excitation energies between \num{1.49} and \SI{2.05}{eV} (top to bottom). Amplitudes of the spectra were normalized to 1; the spectra are plotted vertically offset. Colored stripes and signs (+, $\times$) indicate two main contributions of the \D. For high excitation energies, the Raman signal of POF (bottom spectrum) overlaps with the weak \textit{D}-mode intensity.}
	\label{fig:d-mode}
\end{figure}

\begin{figure}[tb]
	\centering
		\includegraphics[width=\linewidth]{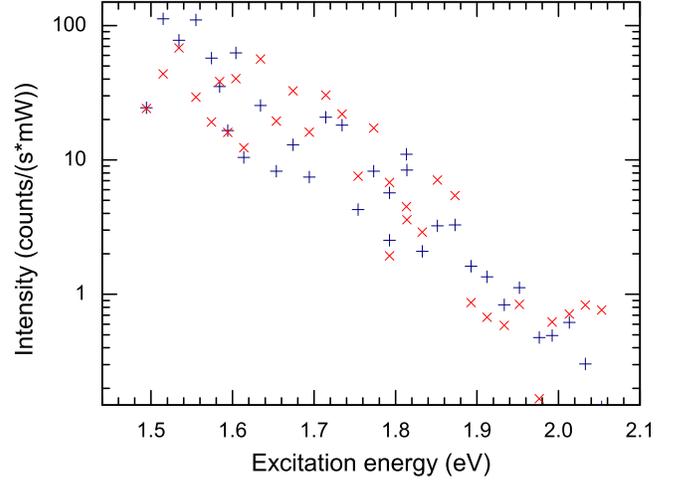}
	\caption{(Color online) Resonance behavior of the \D\ of the (9,7) enriched sample for excitation between \num{1.49} and \SI{2.05}{eV}. The signs + and $\times$ denote the two contributions marked in Fig.~\ref{fig:d-mode}.}
	\label{fig:intensity}
\end{figure}

\begin{figure}[tb]
	\centering
		\includegraphics[width=\linewidth]{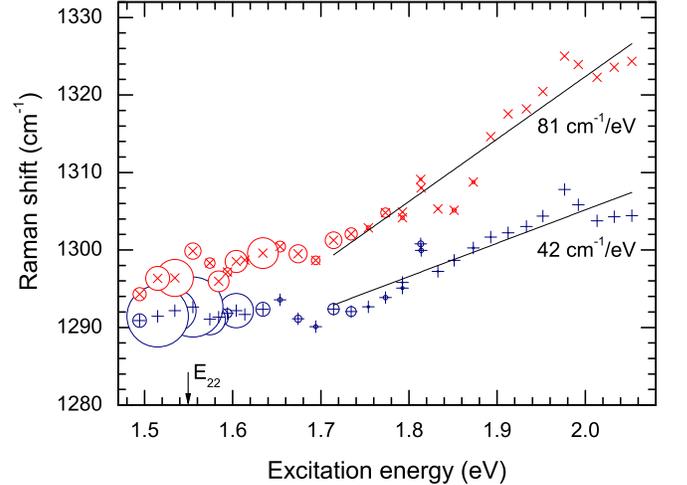}
	\caption{(Color online) Dispersion of the  two contributions of the \D, denoted by the signs + and $\times$ (Fig.~\ref{fig:d-mode}). The radii of the circles indicate the \textit{D}-mode intensities. The arrow indicates the $E_{22}$ transition energy of the (9,7) SWCNT determined from the RBM resonance profile (Fig.~\ref{fig:rbmg}).}
	\label{fig:d-modefit}
\end{figure}

Figure~\ref{fig:d-mode} shows the Raman spectra of the \textit{D}~mode for excitation energies between \num{1.49} and \SI{2.05}{eV}, where the maximum amplitude was normalized to one. In these spectra, the \textit{D}~mode consists of at least two main contributions. They are indicated by two colored stripes and denoted with the signs + and $\times$. For high excitation energies, the Raman signal of POF appears in the spectra due to a decrease of the \textit{D}-mode intensity. The spectra were fitted with two Lorentzians, after the background signal of the POF polymer was subtracted.
\\
Figure~\ref{fig:intensity} shows the measured \textit{D}-mode intensity (peak area) of both contributions (+, $\times$) as function of excitation energy on logarithmic scale. Both contributions show a decrease in intensity by two orders of magnitude for the excitation energy ranging from \num{1.49} to \SI{2.05}{eV}.
\\
The dispersion of the \textit{D}-mode frequency with excitation energy is shown in Fig.~\ref{fig:d-modefit}. The frequencies are represented by + and $\times$ signs, the intensities of the \textit{D} mode contributions are indicated by the radii of the circles.\\
The strongest \textit{D}-mode intensity is observed, when the RBM of the (9,7) and (9,8) SWCNTs is in resonance. Unexpectedly, the well-known dispersion behavior of the \D \ \cite{Maultzsch2001} is only observable above a certain excitation energy (\SI{1.7}{eV} for the SWCNTs in our sample).  Here, the \textit{D}-mode intensity is strongly decreased, see Fig.~\ref{fig:intensity}; the dispersions of the two contributions can be linearly fitted to \SI{81}{cm^{-1}/eV} and \SI{43}{cm^{-1}/eV} (Fig.~\ref{fig:d-modefit}).
\\
We explain our results tentatively by attributing the weakly dispersive, high-intensity appearance of the \D\ (Fig.~\ref{fig:d-modefit}, excitation energy between \num{1.49} and \SI{1.7}{eV}) to an enhanced-dRR scattering very close to the optical transition. 
For the dispersive, low-intensity part of the \D\ (Fig.~\ref{fig:d-modefit}, excitation energy above \SI{1.7}{eV}), we are above the resonance window of the enhanced-dRR scattering, and therefore the intensity decreases. At the same time, the electron bands of SWCNTs exhibit the nearly linear dispersion derived from the nearly linear bands close to the \textit{K}-points of graphite and graphene.
To verify this interpretation, we performed calculation of the resonance behavior of the \D\ for the (9,7) and the (9,8)~SWCNT.

\section{Simulation}       
\subsection{Simulation details}
\begin{figure}[tb]
	\centering
		\includegraphics[width=\linewidth]{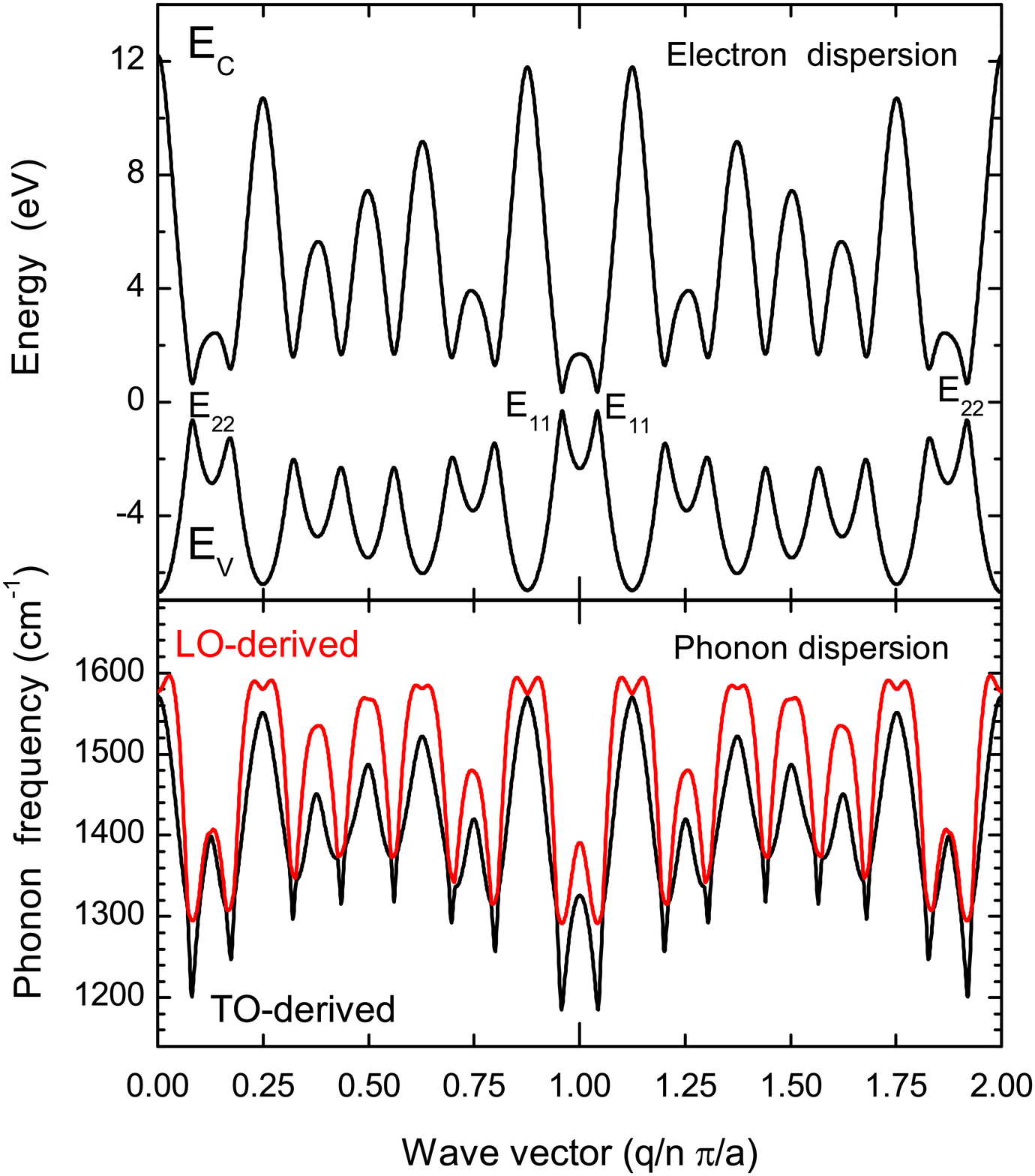}
	\caption{(Color online) Calculated electron and phonon dispersion of the (9,7) SWCNT. The optical transitions $E_{11}$ and $E_{22}$, and the LO- and TO derived phonon branches are indicated. The wave vector is given in units of helical quantum numbers \cite{Damnjanovic2000,reich2008carbon} with $q=\num{386}$, $n=\num{1}$ and lattice translation vector $a=\SI{59.22}{\angstrom}$.}
	\label{fig:disp}
\end{figure}

The \D\ is calculated by a numerical simulation of the dRR scattering. The differential cross section is proportional to $|\mathcal{K}|^2$ \cite{Heitler1954}. The matrix element $\mathcal{K}$ for the scattering with one phonon and one defect is given by~\cite{Maultzsch2001}:
\begin{footnotesize}
\begin{equation*}
\begin{split}
\mathcal{K}=&\sum_{a,b,c}\\\biggl[&\frac{\mathcal{M}_{e-r}\mathcal{M}_{e-def}\mathcal{M}_{e-ph}\mathcal{M}_{e-r}}{(E-E^e_{a}-i\gamma)(E-E^e_{b}-\hbar\omega_{ph}-i\gamma)(E-E^e_{c}-\hbar\omega_{ph}-i\gamma)}\\
&+\frac{\mathcal{M}_{e-r}\mathcal{M}_{e-ph}\mathcal{M}_{e-def}\mathcal{M}_{e-r}}{(E-E^e_{a}-i\gamma)(E-E^e_{b}-i\gamma)(E-E^e_{c}-\hbar\omega_{ph}-i\gamma)}\biggr]
\end{split}
\label{eq:res}
\end{equation*}
\end{footnotesize}
% Klammergrößen \big, \Big, \bigg oder \Bigg 
The first term describes the scattering process by first scattering the electron by a phonon and then the hole by a defect; the second term describes the scattering process by first scattering the electron by a defect and then the hole by a phonon. $E$ is the excitation energy, $E^e_{x} (x=a,b,c)$ are the energy differences between excited and ground state of the electronic bands, $\hbar\omega_{ph}$ is the phonon energy and $\gamma$ is a empirical broadening factor taking into account the lifetime of the intermediate electronic states. The different $\mathcal{M}$'s are the matrix elements for the interaction between electron and radiation ($\mathcal{M}_{e-r}$), phonon ($\mathcal{M}_{e-ph}$), or defect ($\mathcal{M}_{e-def}$). Each $\mathcal{M}$ was set constant for the simulation; $\gamma$ was set to \SI{40}{meV}.
\\ 
The electronic bands were calculated from a third nearest neighbor tight-binding approximation \cite{Reich2002}. The phonon dispersion relations were calculated with a symmetry-based force constants approach adapted to the experimental phonon dispersion of graphite \cite{Mohr2007,Dobardzic2003} with the program POLSym \cite{Milosevic1996}. In chiral SWCNTs, both the longitudinal optical (LO) and the transverse optical (TO) phonons are Raman active at the $\Gamma$ point, in contrast to zigzag or armchair SWCNTs \cite{reich2008carbon}. Inside the Brillouin zone, they lose their defined longitudinal or transverse character; therefore they will be called here LO- and TO-derived phonon branch.
\\ 
Figure~\ref{fig:disp} shows the calculated electron and phonon dispersion of the (9,7)~SWCNT in units of helical quantum numbers \cite{Damnjanovic2000,reich2008carbon}. The transition energies $E_{11}$ and $E_{22}$ in the electron dispersion, and the LO- and TO-derived phonon branches in the phonon dispersion are denoted. In our simulation, $E_{11}=\SI{0.67}{eV}$ and $E_{22}=\SI{1.28}{eV}$ \cite{note1}.

\subsection{Calculated dispersion and resonance behavior of the \D}
\begin{figure}[tb]
	\centering
		\includegraphics[width=\linewidth]{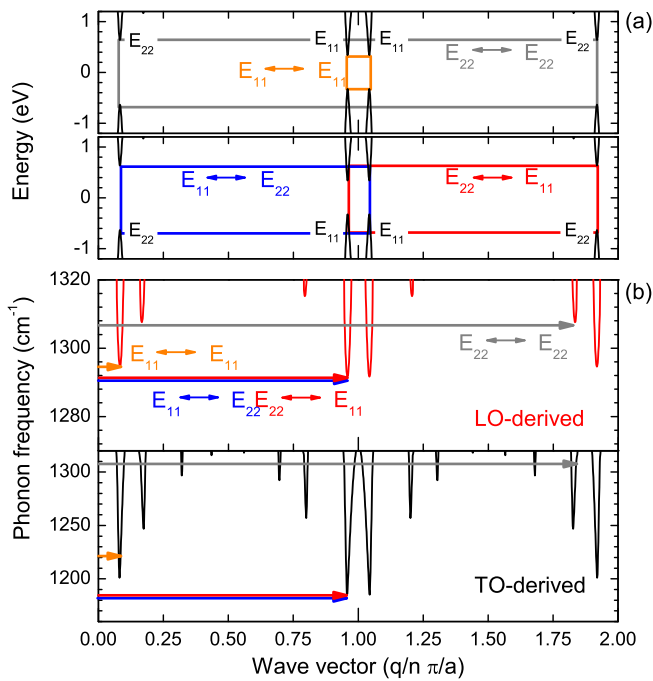}
	\caption{(Color online) (a) Electron dispersion from Fig.~\ref{fig:disp} zoomed to the energy region of the optical transitions $E_{11}$ and $E_{22}$. The upper panel shows a sketch for the "`symmetric"' scattering ($E_{11}\leftrightarrow E_{11}$, $E_{22}\leftrightarrow E_{22}$) of electron and hole; the lower panel for the antisymmetric scattering ( $E_{11}\leftrightarrow E_{22}$, $E_{22}\leftrightarrow E_{11}$). (b) The corresponding phonons for the scattering processes in (a) are drawn into the LO-derived (upper) and the TO-derived branch (lower panel).}
	\label{fig:dispzoom3}
\end{figure}

\begin{figure}[tb]
	\centering
		\includegraphics[width=\linewidth]{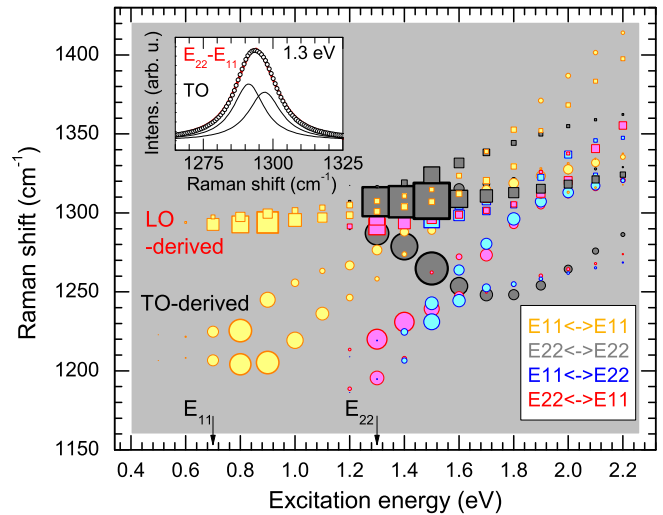}
	\caption{(Color online) Calculation of the different contributions of \D\ for the (9,7) SWCNT. Different colors denote the scattering path; the shapes of the symbols denote the used phonon branch. The symbol size indicates the intensity. The inset shows exemplary a fit of a simulated spectrum with two Lorentzians.}
	\label{fig:calpath}
\end{figure}

\begin{figure}[tb]
	\centering
		\includegraphics[width=\linewidth]{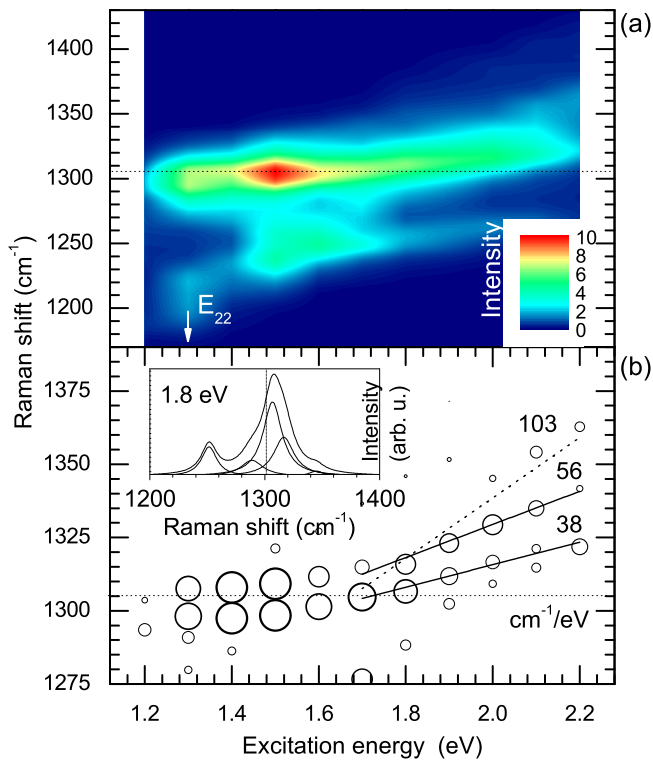}
	\caption{(Color online) (a) Simulated \textit{D}-mode intensity as a color plot with consideration of all interference effects. (b) Dispersion of the simulated Raman spectra, deduced by fitting with Lorentzians (inset). The radii of the circles correspond to the simulated \textit{D}-mode intensity.}
	\label{fig:fitcolor97}
\end{figure}

We performed simulations with excitation energies in the range of both optical transitions $E_{11}$ and $E_{22}$. Therefore, up to four scattering processes for the electrons are in principle possible: $E_{11}\leftrightarrow E_{11}$, $E_{22}\leftrightarrow E_{22}$, $E_{11}\leftrightarrow E_{22}$ and $E_{22}\leftrightarrow E_{11}$. A sketch of these processes for the lowest possible excitation energy is shown in Fig.~\ref{fig:dispzoom3}(a).
Including the two phonon branches and the four scattering processes, eight different processes contribute to our \textit{D}-mode simulation [Fig.~\ref{fig:dispzoom3}(b)].\\ 
In contrast to Ref.~\cite{Maultzsch2001}, where we assumed that the disorder only breaks the pure translational symmetry and not, to first approximation, the rotational symmetry of the nanotube, we now allow the entire phonon branch with helical quantum number $\widetilde{m}=\num{0}$. This corresponds in linear quantum numbers to a change in band index $m$, which would not be allowed if the rotational symmetry was conserved.
\\
To analyze differences between the various contributions of the simulated Raman spectra, we first calculated each scattering process separately, thereby neglecting possible interference effects. Each spectrum of the different contributions was phenomenologically fitted with two Lorentzians (inset Fig.~\ref{fig:calpath}) considering the shorter and longer phonon wave vector from the scattering processes at a given energy.
\\
Fig.~\ref{fig:calpath} shows the fitted data of the simulated \D\ for each of these eight different contributions.  Different colors denote the different scattering processes (color code as in Fig.~\ref{fig:dispzoom3}) and the shapes of the symbols denote the used phonon branches. The symbol size indicates the intensity obtained from the fit. 
\\
From these results, some general observations can be made.
First, all scattering processes show an enhanced intensity near the corresponding optical transitions ($E_{11}$, $E_{22}$). For the symmetric scattering within the same optical transition ($E_{11}\leftrightarrow E_{11}$ and $E_{22}\leftrightarrow E_{22}$), the \textit{D}-mode intensity of the outgoing resonance ($\approx$~\SI{160}{meV} above the optical transition) is always stronger than for the incoming resonance. The reason is that for the incoming resonance, the electron scattered with a phonon will always be below the optical transition (i.e. band extrema with high density of states in our simulation), whereas for the outgoing resonance, scattering with a phonon matches the optical transition.
Second, the \D\ becomes dispersive at energies above the outgoing resonance, as observed in experiments.
\\
A peculiarity in the dispersion is the scattering process for the shorter wave vector of the scattering $E_{22}\leftrightarrow E_{22}$ involving the TO-derived branch  (Fig.~\ref{fig:calpath}, gray circles, lower branch). Here, an unusual negative dispersion can be observed. This is due to the fact that the phonon wave vector $q$ of the lowest possible optical transition does not correspond to the lowest frequency of the TO-derived phonon branch [Fig.~\ref{fig:dispzoom3}(b), lower panel, gray arrow]. For smaller $q$~vectors (corresponding to higher excitation energy), therefore the frequency of the \D\ first drops before it rises again after passing the minimum of the phonon dispersion.
\\
A full simulation of the \textit{D}-mode intensity of the (9,7) SWCNT with an overlap of the eight different scattering processes, necessarily including all interference effects \cite{Maultzsch2004a}, is shown in Fig.~\ref{fig:fitcolor97}(a). An excitation-energy range similar to the experiment was chosen.\\ 
Similar to the experiment, the simulated \D\ shows a strong intensity in a small energy range, followed by an intensity decrease and a starting dispersive behavior for higher energies. A comparison with the calculated data from Fig.~\ref{fig:calpath} shows that the highest intensity belongs to the scattering $E_{22}\leftrightarrow E_{22}$ with the LO-derived phonon branch (gray squares).\\
To deduce the dispersion of the full simulation, the Raman spectra were phenomenologically fitted with five to six Lorentzians [inset of Fig.~\ref{fig:fitcolor97}(b)]. The result is shown in Fig.~\ref{fig:fitcolor97}(b). Here the radii of the circles correspond to the \textit{D}-mode intensities.\\
The inset of Fig.~\ref{fig:fitcolor97}(b) shows that in the frequency range of the experimental data ($\num{1280}-\SI{1330}{cm^{-1}}$), the simulated Raman spectra consist of two main contributions. These two main contributions are due to an overlap of the different scattering processes. For these two main contributions a dispersion of \num{38} and \SI{56}{cm^{-1}/eV} (solid lines) is obtained. 
\\
For our calculation of the \textit{D}-mode resonance of the (9,8) SWCNT (not shown) we found a similar behavior as described above. For that reason and due to the much weaker RBM intensity of the (9,8) SWCNT in comparisons to the (9,7) SWCNT we neglected this contribution of the simulated \D\ in the following comparison between calculation and experimental data. 
\section{Comparison of calculations with experimental data}
\begin{figure}[tb]
	\centering
		\includegraphics[width=\linewidth]{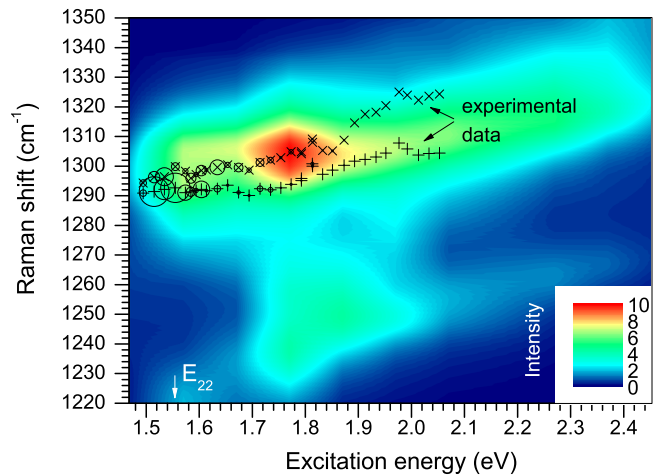}
	\caption{(Color online) Comparison of dispersion behavior between simulations (color intensity plot) and experimental data (symbols).}
	\label{fig:calexpcontour}
\end{figure}

In Fig.~\ref{fig:calexpcontour}, the experimental data (Fig.~\ref{fig:d-modefit}) are plotted into the intensity color plot of our simulations [Fig.~\ref{fig:fitcolor97}(a)]. In our simulations, the energies of the optical transitions were slightly underestimated. For better comparison, the simulations were shifted by $+\SI{0.28}{eV}$ to bring the theoretical optical transition energies to the same energy as in experiment.
\\
In general, the agreement between calculation and experiment is quite good as discussed above: In both cases an enhanced \textit{D}-mode intensity is visible, as well as the non-dispersive parts have similar frequencies. Further, at higher excitation energies, we observe a similar dispersion of the \D.
\\    
The two most obvious differences between simulation and experiment are: First, the \textit{D}-mode intensity in the simulation decreases only linearly by a factor of two and not exponentially by two orders of magnitude; second, the highest \textit{D}-mode intensity in the simulations is shifted to higher excitation energy. This shift is caused by the strong outgoing resonance in the simulation, which we do not observe in the experiment. Both differences might be due to neglecting the wave-vector dependence of matrix elements $\mathcal{M}$ and excitonic effects in our simulations. Also an intensity decrease of the outgoing resonance due to the non-Condon effect (recently observed for the \textit{G}~mode \cite{Duque2011}) might play a role for the \textit{D}-mode intensity and was not taken into account in our simulation.
\\
\begin{figure}[tb]
	\centering
		\includegraphics[width=\linewidth]{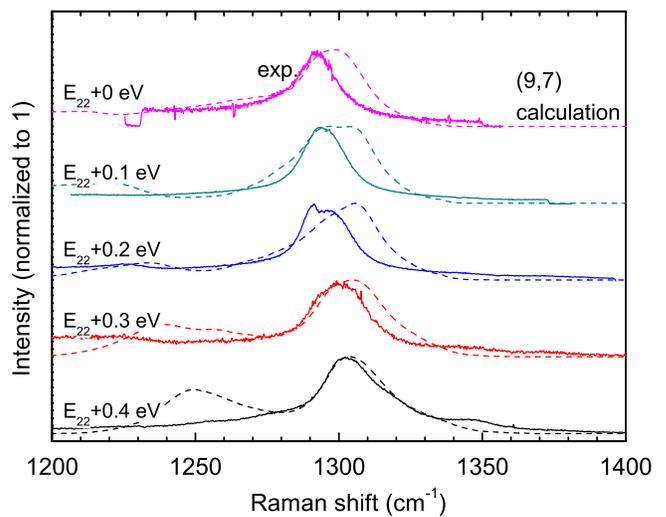}
	\caption{(Color online) Comparison of calculated (dashed lines)and experimental (solid lines) Raman spectra. Amplitudes of the spectra were normalized to 1; the spectra are plotted vertically offset.}
	\label{fig:calexpline}
\end{figure} 
A comparison of the Raman spectra between experiment and simulation is shown in Fig.~\ref{fig:calexpline}. For excitation energies near the optical transition $E_{22}$, a small difference between experiment and simulation is seen. This frequency difference might be caused by inaccuracies in the calculated phonon dispersion, in particular near the wave vectors derivated from the $K$ point in graphene, which contribute most at excitation directly at the optical transition.
Since also the different contributions of the \D\ have slightly different frequencies (Fig.~\ref{fig:calpath}), the observed frequency deviations might also be caused by an overestimation of one of these contributions, again because of neglecting the matrix elements.
\\
Furthermore, an additional peak appears for higher excitation energies in the range of low frequencies in our simulations (Fig.~\ref{fig:calexpline}, at $\approx\SI{1250}{cm^{-1}}$). This contribution corresponds to a scattering processes involving the TO-derived phonon branch (Fig.~\ref{fig:calpath}, circles). Due to the anti-crossing of the phonon bands near the \textit{K} point, the frequencies of the TO-derived phonon branch corresponds to the LO-phonon branch at the $\Gamma$-point for the (9,7) SWCNT. % (and the \textit{K}-point of graphene). 
For $\Gamma$-point scattering (RBM and the \textit{G}~mode) it was shown in calculations that for SWCNTs with a large chiral angle [as for the (9,7) nanotube], the Raman intensity related to the LO-phonon branch is much weaker than for the TO-phonon branch \cite{Yin2007a}.
Therefore we think that the Raman intensity for scattering involving the TO-derived phonon branch at the K point (Fig.~\ref{fig:calexpline}, at $\approx\SI{1250}{cm^{-1}}$) might be much smaller than in our simulations; as we observe in our experiments.     
\\
The unexpected high dispersion of \SI{81}{cm^{-1}/eV} in the experimental data (Fig.~\ref{fig:d-modefit}) might be explained with a jump to contributions from another scattering process as seen in the calculations in Fig.~\ref{fig:fitcolor97}(b), indicated with a dashed line.
Another explanation might be that due to the exponential decrease of the \D\ the still remaining metallic nanotubes contribute to the \D\ for excitation energies above \SI{1.9}{eV}. This would lead to an additional contribution to the \D\ at higher frequencies  \cite {Laudenbach2012} and therefore to an overestimated dispersion.

\section{Conclusion}
In conclusion, we showed the resonance behavior of the Raman \D\ in a sample enriched with (9,7) SWCNTs, thereby following the \D\ of very few, chiral-index identified nanotubes across an optical transition. The \D\ intensity is strongest at the optical transition of the (9,7) SWCNT; the \textit{D}-mode frequency, however, shows almost no dispersion with excitation energy near the resonance. The well-known dispersion of the \textit{D}-mode frequency is observed only above resonance, where the intensity has dropped by approximately one order of magnitude. These results are understood based on numerical simulations of the \textit{D}-mode spectra, although the intensity drop observed in experiment is even stronger than in the simulations. Further calculations therefore should include excitonic effects.
\\
Based on these observations, we expect that the \textit{D}-mode dispersion measured in nanotube ensembles not just reflects the electron and phonon bands of a given nanotube but also a diameter and chiral-angle dependence of the \textit{D}-mode frequency \cite{Kuerti2002,SouzaFilho2002}.    
\\
Our results further show that the \textit{D}-mode intensity does not exclusively depend on the defect density of the nanotubes but also on the excitation energy. This should be considered together with the excitation-energy dependence of the \textit{G}-mode intensity \cite{Fouquet2009} when taking the \textit{D}/\textit{G} intensity ratio as a measure for the defect density.  

\textbf{Acknowledgment.}
We thank M. Damnjanovi\'{c}, I. Milo{\v s}evi\'{c} and their group (University of Belgrade, Serbia) for helpful discussions and for providing the POLSym code. This work was supported  by the European Research Council (ERC) under grant no. 259286.

\end{document}